\documentclass[10pt,twocolumn,aps,prb,showpacs,superscriptaddress]{revtex4}
\usepackage[T1]{fontenc}
\usepackage[latin9]{inputenc}
\usepackage{textcomp}
\usepackage{amsmath}
\usepackage{amssymb}
\usepackage{graphicx}

\makeatletter
\@ifundefined{textcolor}{}
{%
 \definecolor{BLACK}{gray}{0}
 \definecolor{WHITE}{gray}{1}
 \definecolor{RED}{rgb}{1,0,0}
 \definecolor{GREEN}{rgb}{0,1,0}
 \definecolor{BLUE}{rgb}{0,0,1}
 \definecolor{CYAN}{cmyk}{1,0,0,0}
 \definecolor{MAGENTA}{cmyk}{0,1,0,0}
 \definecolor{YELLOW}{cmyk}{0,0,1,0}
}

\usepackage{bm}
\usepackage{color}

\makeatother

\begin{document}

\title{Two components for one resistivity in LaVO$_{3}$/SrTiO$_{3}$ heterostructure.}

\author{H.~Rotella}
\affiliation{Laboratoire CRISMAT, CNRS UMR 6508, ENSICAEN et Universit$\acute{e}$
de Caen, 6 Bd Mar$\acute{e}$chal Juin, 14050 Caen Cedex 4, France.}
\affiliation{NUSNNI-NanoCore, National University of Singapore, Singapore 117411.}

\author{O.~Copie}
\affiliation{Laboratoire CRISMAT, CNRS UMR 6508, ENSICAEN et Universit$\acute{e}$
de Caen, 6 Bd Mar$\acute{e}$chal Juin, 14050 Caen Cedex 4, France.}
\affiliation{CEA, DSM/IRAMIS/SPEC, F-91191 Gif-sur-Yvette Cedex, France.}

\author{A.~Pautrat}
\email{alain.pautrat@ensicaen.fr}
\thanks{Corresponding author}
\affiliation{Laboratoire CRISMAT, CNRS UMR 6508, ENSICAEN et Universit$\acute{e}$
de Caen, 6 Bd Mar$\acute{e}$chal Juin, 14050 Caen Cedex 4, France.}

\author{P.~Boullay}
\affiliation{Laboratoire CRISMAT, CNRS UMR 6508, ENSICAEN et Universit$\acute{e}$
de Caen, 6 Bd Mar$\acute{e}$chal Juin, 14050 Caen Cedex 4, France.}

\author{A.~David}
\affiliation{Laboratoire CRISMAT, CNRS UMR 6508, ENSICAEN et Universit$\acute{e}$
de Caen, 6 Bd Mar$\acute{e}$chal Juin, 14050 Caen Cedex 4, France.}

\author{D.~Pelloquin}
\affiliation{Laboratoire CRISMAT, CNRS UMR 6508, ENSICAEN et Universit$\acute{e}$
de Caen, 6 Bd Mar$\acute{e}$chal Juin, 14050 Caen Cedex 4, France.}

\author{C.~Labb$\acute{e}$}
\affiliation{Laboratoire CIMAP, CNRS UMR 6252, CEA, ENSICAEN et Universit$\acute{e}$
de Caen, 6 Bd Mar$\acute{e}$chal Juin, 14050 Caen Cedex 4, France.}

\author{C.~Frilay}
\affiliation{Laboratoire CIMAP, CNRS UMR 6252, CEA, ENSICAEN et Universit$\acute{e}$
de Caen, 6 Bd Mar$\acute{e}$chal Juin, 14050 Caen Cedex 4, France.}

\author{W.~Prellier}
\affiliation{Laboratoire CRISMAT, CNRS UMR 6508, ENSICAEN et Universit$\acute{e}$
de Caen, 6 Bd Mar$\acute{e}$chal Juin, 14050 Caen Cedex 4, France.}

\date{\today}
\begin{abstract}
A series of 100 nm LaVO$_{3}$ thin films  have been synthesized on (001)-oriented
SrTiO$_{3}$ substrates using the pulsed laser deposition technique, and the effects of growth temperature are analyzed.
Transport properties reveal a large electronic mobility
and a non-linear Hall effect at low temperature. In addition,
a cross-over from a semiconducting state at high-temperature to a
metallic state at low-temperature is observed, with a clear enhancement
of the metallic character as the growth temperature increases. Optical absorption measurements combined with the two-bands analysis
of the Hall effect show that the metallicity is induced by the diffusion of oxygen vacancies in
the SrTiO$_{3}$ substrate. These results allow to understand that the film/substrate heterostructure behaves as an original semiconducting-metallic 
parallel resistor, and electronic transport properties are consistently explained.
\end{abstract}


\maketitle
\subsection{Introduction.}
The discovery of unexpected conducting or superconducting behaviors when stacking
 two band insulators like LaAlO$_{3}$ (LAO) and SrTiO$_{3}$
(STO) has motivated intense activities to understand these
phenomena \cite{ohtomo04,reyren07,caviglia08}. The origin of the
highly mobile charge carriers appears the central point of the debate. Interpretations are based on an electronic reconstruction at the interface due to a polar discontinuity between the two insulators \cite{nakagawa06,thiel06,reyren07,caviglia08}, on cation intermixing or on oxygen vacancies \cite{herranz06,kalabukhov07,siemons07}.
It has also been shown that 2D electronic conduction can be observed
for peculiar controlled conditions. In particular, the oxygen pressure
used during the film growth and the annealing conditions are critical parameters. 
They  allow to tune the amount of oxygen vacancies and the dimensionality of transport
properties \cite{basletic08,copie09}. Using a Mott insulator having
a strong coulomb interaction instead of a band insulator should allow
to take advantage of the rich properties of electronic correlated
systems \cite{okamoto}, and other systems such as LaTiO$_{3}$/SrTiO$_{3}$
(LTO/STO) have been investigated \cite{ohtomo02,biscaras10}. Similarly
to LAO/STO, LTO/STO displays a conducting behavior which was explained by
an electronic reconstruction, and superconductivity emerges at low temperature \cite{biscaras10}.
LaVO$_3$/SrTiO$_3$ (LVO/STO)  is another interesting oxide heterostructure, LVO being a Mott insulator that exhibits an antiferromagnetic spin order \cite{mahajan92}.  Besides the existence of a 2D electron gaz proposed at the LVO/STO interface \cite{hotta07},  new phenomena could arise from the combination of confined carriers at the interface and strong correlations in LVO. Thin
films of LVO/STO were synthesized using the pulsed laser deposition (PLD) technique by Hotta
\emph{et al.} \cite{hotta06}. Importantly, a pure LVO phase can be
formed only under low partial pressure of oxygen (typ. < 10$^{-5}$
Torr), and a post-annealing under oxygen should be avoided since the oxidized
form of LaVO$_{4}$ is quickly stabilized. Therefore, the problematic of oxygen vacancies is specially relevant for
LVO/STO. Regarding the physical properties, this heterostructure
presents a conducting behavior and a low temperature non linear Hall
effect, both explained by the intrinsic mechanism of electronic
reconstruction at the interface \cite{hotta07}. The primary role of
interface effects was also recently proposed to explain the LVO/STO metallic properties \cite{he},  and the presence of oxygen vacancies
was not considered as a single STO substrate placed in the deposition
chamber was not conducting after the ablation process.
However, this experiment is not sufficient to rule out the formation of oxygen
vacancies in a heterostructure. For example in LAO/STO,
the presence of oxygen vacancies in STO is strongly enhanced by the
presence of a LAO epilayer due to the difference of activation
energy for oxygen diffusion between the two materials \cite{kalabukhov07}.
The process of doping is then mainly driven by oxygen transfer between
an oxygen deficient growing film and its substrate \cite{schneider10}.
Similar argument could \textit{a priori} be applied to the LVO/STO
case \cite{cherry95}.  Recently, the LVO/STO heterostructure was shown to be a potential efficient solar cell
 by Assmann $\textit{et al}$ \cite{Assmann}.
In this approach, the presence of an effective electric field due to a polar discontinuity at the interface between
 the two material is an essential ingredient. It is then important to have a better understanding of the conducting mechanisms in this system.

Here, we address how the growth
conditions affect the transport properties in a series of LVO/STO films. Since the pressure can not be significantly changed
to stabilize V$^{3+}$ in LVO, the growth temperature is the most accurate parameter to be considered. We have synthesized
a series of high quality LVO thin films deposited on (001)-oriented STO
substrates at different growth temperatures, and investigated both
the structural and magnetotransport properties. We show that the transport properties are dominated by a semiconducting-metallic
competition which is sensitive to the growth temperature. This is
explained by considering the contribution of the STO substrate which is 
doped by thermally activated oxygen vacancies. This appears to be
a generic process in STO based heterostructure.

\subsection{Experimental.}
Epitaxial LaVO$_{3}$ (LVO) thin films were prepared by the pulsed
laser deposition technique on (001)-oriented SrTiO$_{3}$ (STO) substrates
(cubic with $a$=3.905 \AA ). We have used as-received substrates: No
 specific treatment has been 
performed to select the SrO or TiO$_2$ substrate termination.
Note that as-received substrates are TiO$_2$ terminated \cite{STO1,STO2}.
 A KrF laser ($\lambda=248nm$)
with a repetition rate of 2 Hz was focused onto a LaVO$_{4}$ polycrystalline
target at a fluence of $\simeq$ 2 J/cm$^{2}$. The substrate was
at a temperature ranging from 600-750$^{\circ}{C}$ under a dynamic
vacuum around $10^{-5}$ mbar (base pressure). The target-substrate distance was fixed
at 8.5 cm. The number of pulses was adjusted to obtain the desired
thickness of 100 nm \cite{rotella12}. The X-ray diffraction
measurements were performed with a $\theta$/2$\theta$ diffractometer
Seifert XRD 3000P ($\lambda_{Cu}$=1.5406 \AA ) for room temperature
measurements and with a $\theta$/2$\theta$ X'pert Pro MPD PANalytical
($\lambda_{Co}$=1.789 \AA ) for measurements at different temperatures.
LVO is orthorhombic ($Pnma$ (\#62)) in its bulk form at room temperature
and it undergoes a structural transition below 140K into a monoclinic
($P2_{1}/c$ (\#11)) symmetry \cite{bordet93}. Previously, we have
shown that similar LVO thin film grown at 700$^{\circ}{C}$ presents
a distorted orthorhombic structure due to the compressive stress induced
by STO, with original VO$_{6}$ rotations \cite{rotella12}. Transport
and galvanomagnetic properties were measured using a Quantum Design
PPMS. Electrical contacts were realized
using Al/Au wire-bonding. Sheet resistivity is defined as
R$_{s}$=$\rho$/t, $\rho$ being the bulk resistivity and $t$ the thickness.
The absorbance spectra were obtained using a Perkin Elmer spectrophotometer
Lambda 1050. 

\subsection{Results and discussion.}

Figure 1a depicts the $\theta$-2$\theta$ X-ray diffracted pattern
of the LVO films grown at various temperatures.
The thickness was first defined by the number of laser pulses and was checked to be $t$ =100 nm  for all the series by fitting the diffracted patterns using the DIFFaX software \cite{difax}.
The sharpness of the peaks and the presence of Laue fringes confirm
the high structural quality and a low interface roughness in agreement with previous reports \cite{boullay11,tan13}.
As the growth temperature decreases, the peak associated to LVO shifts
towards smaller 2$\theta$ angle, indicating an increase of the out-of-plane parameter (Fig.1b).
From reciprocal space mapping, we know that our LVO films are coherently in plane- strained all over the thickness \cite{rotella12}, and the expanded c-axis corresponds to an increase of the volume.
 Such expansion of the c-axis parameter can be induced 
by oxygen vacancies during the film deposition as seen previously in copper oxides films \cite{caxis}, STO/STO homoepitaxial films \cite{othomo07} and recently observed in PrVO$_{3}$
films grown on STO \cite{copie09bis}.

The sheet resistance R$_s$ of the films was measured as a function of the temperature
$T$ (See Fig.2). LVO is a localized system with semiconducting conductivity
\cite{rogers65} even in the presence of some cationic and/or oxygen vacancies
\cite{hur94,jamali11} while pure stoichiometric STO is insulating. However, 
a clear metallic behavior with (dR/dT)>0 is clearly observed up
to a temperature T$^{*}$. For T>T$^{*}$, dR/dT<0 indicates a semiconducting-like
behavior. T$^{*}$ varies from samples to samples with different temperature
growth. The room temperature value of
the sheet resistance R$_{s}$ varies from R$_{s}$ $\sim2.71$ to $28.5 k\Omega/\square$
with a minimum $\sim 100\Omega/\square$  at low temperature
for the most conducting sample. These values  are in good agreement with those reported in other heterostructures (LTO/STO \cite{biscaras10},
LAO/STO \cite{caviglia08}),  where low dimensional conduction and superconductivity
have been reported. They are also close to values measured in LVO/STO \cite{hotta07,he}.
An upturn of the resistance at low temperatures was noted by Hotta
\textit{et al.} and explained by an incipient localization mechanism \cite{hotta07}.
 Here, this effect is not seen in our measurements. This is possibly due to
a smaller amount of defects (such as grain boundaries) in agreement with the lowest residual
resistivity ($\sim 100\Omega/\square$ for the most conducting film).

To understand the origin of the change in resistivity at T$^{*}$, we
focus on the film grown at 700$^{\circ}$C (Fig.3) with T$^{*}\sim225K$.
Since the film has a distorted orthorhombic structure with significant
tilts of VO$_{6}$ octahedra at room temperature,\cite{rotella12}
it is worth asking wether the thermal variation of transport properties
correlates with changes in lattice parameters and/or structural distortion \cite{adrian} .
In particular, it is not known if the orthorhombic
to monoclinic structural phase transition observed at 140 K in bulk samples \cite{bordet93} is
still observed in the LVO thin film. Note, however that such transition
favors orbital ordering, and a more localized state at low temperature,
\cite{goodenough} which is hardly compatible with the metallicity observed
here. The evolution of a-axis and c-axis lattice parameters of the film
were recorded as function of temperature under cooling conditions (Fig.
3b and 3c). While a small curvature is observed over a broad temperature range for the c-axis parameter,
there is no discontinuous changes that could indicate a transition,
and the temperature of c-axis minimum is shifted by 60K lower than
T$^{*}$(Fig. 3b). We conclude that the apparent semiconducting-metal
transition is unrelated to the structural evolution of the LVO film.
To go further, we have investigated in details the galvanomagnetic
properties of the same sample. The magnetoresistance
(MR) and Hall effect are measured at different fixed temperatures. As shown in
Fig. 4, a positive MR is observed at low temperature which reaches
MR(14T)=$\frac{R(14T)-R(0T)}{R(0T)}\sim$ 0.01 at 10K. MR increases
continuously when decreasing the temperature, as expected for a metallic
MR driven by mean free path effects. This increase is more pronounced
for T$\lesssim$ 100K, an effect that will be discussed hereafter. Note that
a drastic change of MR can be expected at a metal-to-insulator transition
due to Fermi surface reconstruction. Here, we observe that the sign of MR
changes at high temperature, but only continuously, and at
a temperature larger than $T^{*}$ (see inset of Fig. 4). This confirms
that $T^{*}$ corresponds to a smooth cross-over between two conducting
regimes rather than to a genuine transition.

We have measured the Hall resistance R$_{xy}$ at various temperatures. Its sign is typical of n-type carriers.
 A  preliminary analysis was made by fitting the data in the low field regime (i.e. where R$_{xy}$ is a linear
function of the applied field) using a single-band model. The mobility $\mu$ and the sheet carriers
density $n_{s}$  have been calculated this way  as a function of the temperature
and the results are reported in Fig. 5. We found $\mu\approx$1.3
10$^{3}$ $cm^{2}$/V.s and $n_{s}\approx$ 2.10$^{13}$ /$cm^{2}$
at low temperature. The electronic mobility is consistent with the
high values reported in other STO based heterostructures, including
LVO/STO deposited under low pressure, but are also observed in single phase STO doped by oxygen vacancies,
as already discussed for the LAO/STO case \cite{siemons07,herranz06}.

When looking closely at the R$_{xy}$(B) curves, a non-linear field
variation of the Hall resistance is evidenced. The non-linearity
is more important for T $\lesssim$ 100 K. A non-linear Hall resistance
was observed by Hotta \emph{et al.} in conducting LVO/STO at low temperature,
\cite{hotta07} and was proposed to arise from spin effects such
as antiferromagnetic fluctuations at the interface \cite{hotta07}
or strong spin-orbit coupling \cite{weng10}. In other heterostructures,
a similar non-linear effect was analyzed with a conventional galvanomagnetic
coupling, using a two bands rather than a single band model (in LTO/ST \cite{kim10,biscaras12},
in LAO/STO\cite{bell09}).

In a two-bands model, two groups of charge carriers are
considered with mobilities
$\mu_{1}$ and $\mu_{2}$ and densities $n_{1}$ and $n_{2}$, respectively. In this case, the Hall coefficient $R_{H}=\frac{R_{xy}}{B}$
can be written as \cite{ziman}:

\begin{equation}
R_{H}=1/e\frac{n_{1}\mu_{1}^{2}+n_{2}\mu_{2}^{2}+(n_{1}+n_{2})\mu_{1}^{2}\mu_{2}^{2}B^{2}}
{(n_{1}\mu_{1}+n_{2}\mu_{2})^{2}+(n_{1}+n_{2})^{2}\mu_{1}^{2}\mu_{2}^{2}B^{2}}
\end{equation}

Due to the four unknown parameters, it is necessary to constraint
the fit. We rewrite equation (1) using the approach used in \cite{arushanov,Laiho} :

\begin{equation}
R_{H}=\frac{R_{0}+R_{\infty}\mu_{*}^{2}B^{2}}{1+\mu_{*}^{2}B^{2}}
\end{equation}

which has two fitting parameters $\mu^{*}$ and $R_{\infty}$, $R_{0}$
being the zero field limit of $R_{H}$ which is graphically determined.
With the value of the zero field resistivity $R_{xx}=(e.n_{1}.mu_{1}+e.n_{2}.\mu_{2})^{-1}$,
the following equations give four independent parameters $\mu_{1}$,
$\mu_{2}$, $n_{1}$ and $n_{2}$

\begin{equation}
\begin{split}A=(R_{0}/R_{xx}+\mu_{*})/2\\
\mu_{1}=A+(A^{2}-\mu_{*}R_{\infty}/R_{xx})^{1/2}\\
\mu_{2}=A-(A^{2}-\mu_{*}R_{\infty}/R_{xx})^{1/2}\\
C=(\mu_{1}/\mu_{2})(\mu_{*}-\mu_{2})/(\mu_{1}-\mu_{*})\\
n_{1}=\frac{1}{eR_{\infty}(1+C)}\\
n_{2}=\frac{C}{eR_{\infty}(1+C)}
\end{split}
\end{equation}

Analysis of the data using this two-bands model is very satisfactory,
as demonstrated by the agreement between the experimental data and the modeled curve reported in the Fig.6.
The extracted parameters are shown in the Fig.7 for all temperatures. At the lowest temperature, we find carriers densities ratio $n_{1}/n_{2}\approx$
0.14 and mobilities ratio $\mu_{1}/\mu_{2}\approx$ 3.2. One open
question is whether the two-bands scenario arises from spatially separated
carriers with one of the carriers family located at the interface
of the heterostructure \cite{biscaras12}. Another possibility is
that these different carriers are intrinsic to STO due to its complex
electronic band structure with at least the presence of one heavy
and one light electron band \cite{mattheis72,ahrens07,lin13}. The pertinence
of a multiband scenario intrinsic to STO was recently confirmed
by Laukhin \textit{et al} based on measurements and analysis of the
pressure dependence of electronic mobilities and carrier density \cite{Laukhin12}, and by Lin \textit{et al} \cite{lin13}.
From literature, the ratio of effective
mass $\mathit{m_{1}/m_{2}}$ is expected $\sim$ 10-20 at low temperature \cite{santander}. The mobilities
are related to the effective masses by $(\mu_{2}/\mu_{1})\sim(m_{1}/m_{2})^{q}$
with q a constant dependent on the scattering mechanism \cite{ziman}.
Thus, for a dominant charged impurity scattering at low temperature where q=1/2, one finds $(\mu_{2}/\mu_{1})\sim$ 3.1-4.4 in very good agreement with
our result. From magnetoresistance and Hall measurements, a ratio
$\mu_{1}/\mu_{2} \approx$ 2.7 was reported in the low temperature
state of n-type STO \cite{Kuchar}. This value is also close to our result. We finally note that both the MR and the non linearity of Hall effect are reinforced for T <100K. 
This temperature is close to the temperature of the cubic to tetragonal transition of STO, where some changes in its galvanomagnetic properties can be expected \cite{Kuchar}.
 To conclude this analysis, the contribution of STO itself is sufficient to explain the non-linear
 Hall effect in LVO/STO, and we propose that this result can be expanded to 
the case of other STO-based heterostructures.

Since the measure of optical properties can reveal the presence of oxygen
vacancies in STO \cite{optique}, we have measured the optical absorption
of the sample grown at 700$^{\circ}$C, and of an as-received STO
substrate for a reference. As shown in Fig. 8, the absorption
spectra of the (film+substrate) structure shows a broad peak centered
around 2.4 eV, and a smaller one around 1.7 eV, whereas both peaks are absent
in the virgin substrate. These peaks lying below the 3.2 eV band-gap
are characteristic of reduced STO bulk crystals, and are systematically
attributed to the presence of F centers caused by oxygen vacancies \cite{optique}.

Based on the experimental data that evidence the presence of oxygen vacancies
in STO, we propose a simple model to quantify this effect and to explain
the macroscopic transport properties including the apparent metal-
to-semiconducting transition. Indeed, the
low pressure ($\sim$ $10^{-5}$ mbar) used during the LVO growth
induces oxygen vacancies in the deposited layers. These vacancies
prefer to migrate inside the STO due to its low activation energy for vacancies
diffusion \cite{kalabukhov07,cherry95}. The kinetics of this process is thermally
activated, and the relatively high growth temperature (T$_{Growth}$ > 600$^{\circ}$C) allows for their effective migration
during the ablation process. Also, the c-axis parameter is known to expand
with the oxygen vacancies (c-axis expansion effect) \cite{caxis,copie09bis}. If a high-temperature growth favors
oxygen vacancies in the substrate, the deposited LVO film is correlatively
more close to its stoichiometry in oxygen. Consistently, we observe a decrease of the c-axis LVO value when the growth temperature increases.
 Small deviation from perfect La/V
ratio can be favored by oxygen deficiency in LVO \cite{rogers65,seim}.
However, such non-stoichiometry has only a small effect on the transport
properties of LVO which is a robust semiconductor with only small
band gap changes \cite{jamali11}. On the contrary, it is well known that STO becomes conducting 
even with a tiny doping by oxygen vacancies \cite{siemons07,copie09}.  We have then to deal with two
resistors in parallel: the LVO film expected to be semiconducting-like in one side, and the STO substrate being partially conducting on the other side. One
assumption is that the conducting thickness of the STO is fixed
by the diffusion length of oxygen vacancies at the macroscopic scale,
and could be strongly dependent on the growth temperature. Note
this parallel resistor with opposite temperature variation mimics
an equivalent resistor with non-monotonic temperature variation which
presents a resistance maximum at some intermediate temperature \cite{clara},
in qualitative agreement with our measurements (Fig.2). To be more
quantitative, $G_{i}$, $\rho_{i}$ and $t_{i}$ will stand for conductance,
resistivity and thickness, respectively, while i=1,2 are the LVO and STO.
For a parallel resistor, the equivalent conductance G is:

\begin{equation}
G=\frac{1}{R}=G_{1}+G_{2}=\frac{1}{R_{1}}+\frac{1}{R_{2}}=\frac{t_{1}W}{d\rho_{1}}+\frac{t_{2}W}{d\rho_{2}}
\end{equation}

While $t_{1}=100nm$ is the physical thickness of LVO, $t_{2}$ is
an effective thickness corresponding to the conducting part of STO
which is related to the oxygen vacancies diffusion length $\ell$, d and W are respectively the distances between the voltage pads and the film width.
  For all the samples, the low temperature resistance $R(T)$
can be fitted by $R_{res}+A.T^{2}$ over a large temperature range  ($R_{res}$ is the residual resistance). This is typical of a doped-STO contribution
(Fig. 9). This indicates that the conducting paths are short-circuited
by the substrate at low temperature ($G_{1}\ll G_{2}$) and $R\sim R_{2}$
for T $\lesssim$ 100K. Assuming that $R_{2}(T)=R_{0}+A.T^{2}$ is
fulfilled up to 300 K as in bulk STO, $R_{1}(T)$ and $\rho_{1}(T)$
are directly deduced using equation (4).

The bulk LVO conduction is semiconducting-like with a resistivity described
by a thermally activated law: $\rho(T)\propto\exp(E_{a}/KT)$ (where $E_{a}$
is the activation energy, $K$ the Boltzmann constant). As shown in Fig.10,
$R_{1}(T)$ presents such an activated behavior for all the samples.
The extracted energies $E_{a}$ are close to 0.2 eV, a
very reasonable value for conducting processes in rare earth vanadates
such as LVO \cite{rogers65}. The $\rho_{1}$ value taken at room temperature is
also close the resistivity of bulk LVO, i.e. in a range 0.1-0.3 $\varOmega.cm$,
\cite{rogers65} and increases when T$_{growth}$ decreases (Fig.10).
Since the highest growth temperature favors the diffusion of oxygen
vacancies in STO, their amount in LVO is expected to be larger at
the lowest temperature. The presence of oxygen vacancies or charge
defects often reduce the resistance by doping and/or by creating mixed
valencies in oxides. In LVO, we observe an increase of resistance
when we expect the largest oxygen non-stoichiometry. There is however no contradiction with the
literature since the introduction of mixed valency in LVO by non stoichiometry seems to favor the carriers localization with the creation of defect states \cite{jamali11}.
Finally, we conclude that R$_{1}$ has to be associated with the LVO film.

Let us now return to the behavior of STO. The oxygen vacancies diffusion
mechanism is a thermally activated process, with a diffusion constant
given by an Arrhenius dependence $D(T)=D_{0}\exp(-U/KT)$ where $D_{0}$
is the diffusion constant, $U$ the activation energy. Thus, during
a growth time $\tau_{d}$ and at a certain $T_{growth}$ temperature, the
diffusion length in a 1D diffusion is given by $\ell(T_{growth})=2\sqrt{\times D(T_{growth})\times\tau_{d}}$.
Dealing with the room temperature values of conductivity, we neglect
here the possible spreading of carriers due to the unusual large dielectric
constant of STO at low temperature \cite{siemons07,copie09}, and we approximate the effective thickness of
conducting STO by $t_{2}\sim\ell$. At room temperature and below,
we assume that the thermal energy (typ. 0.026 eV at 300 K) is
low enough to neglect any supplementary vacancies diffusion after the growth process. Finally,
the STO conductance $G_{2}$ can be written as: 
\begin{equation}
G_{2}=1/R_{2}\sim2\sqrt{\times D_{0}exp(U/KT_{growth})\tau_{d}}/\rho_{2}
\end{equation}
Here, the important parameter is U, the activation energy for the diffusion
of oxygen vacancies in STO, which is experimentally given by the slope
of $\ln(G_{2})$ as function of $1/T_{growth}$  (Fig. 10). The fit
leads to $U\sim0.9$ eV, in good agreement with literature ($U\sim0.7-1.4$
eV) \cite{oxygen,Desouza}.

In conclusion, we have studied the electronic conduction properties of a
series of LaVO$_{3}$ (LVO) thin films grown at various temperature on SrTiO$_{3}$ (STO) substrates. A pronounced
metallic character is observed as in other heterostructures based on STO.  The better metallicity is observed for the films synthetised at the highest  growth temperature.
 We show, thanks to a detailed analysis of galvanomagnetic and optical properties in one film, that the 
metallic component arises from the SrTiO$_3$ substrate which was populated by oxygen vacancies during the film growth.
 As a consequence, the heterostructure can be thought as a parallel resistor with a semiconducting LVO component and
a metallic STO component, and the equivalent resistance mimics an apparent metal to semiconducting-like transition.
 Even though electronic reconstruction effects
might be present in the system, we conclude that the macroscopic electronic transport
properties are mainly driven by the competition between these two
bulk components. At low temperature, STO dominates the galvanomagnetic properties, including the non-linear Hall effect. 

We thank Fabien Veillon (CRISMAT) and Jerome Lecourt (CRISMAT) for the precious help in respectively
 the physical properties measurements, and the ceramic synthesis for thin film growth. Discussions with and encouragement
 of the emeritus Prof B. Mercey (CRISMAT) were highly appreciated.
We thank also Jean Louis Doualan (CIMAP) for technical help for the optical measurements.
 Partial support of the French Agence Nationale de la Recherche (ANR),
through the program Investissements d'Avenir (ANR-10-LABEX-09-01), LabEx EMC3, the C'Nano program and
the Interreg IVA MEET project are also acknowledged.

\clearpage

\begin{figure}[ht]
\centering \includegraphics{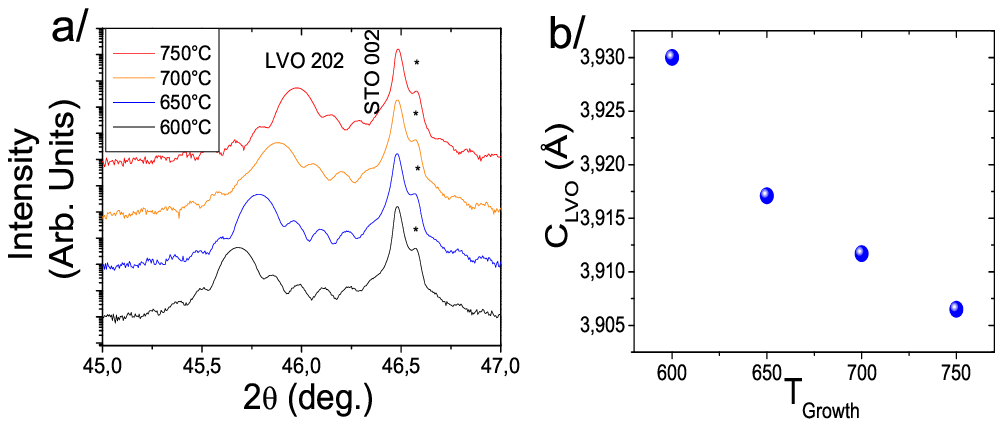} 
\caption{a/ $\theta$-2$\theta$ X-ray diffracted patterns of 100 nm thick films
grown at 600, 650, 700 and 750$^{\circ}$C. The star indicates the
K$_{\alpha2}$ contribution of the (002)SrTiO$_{3}$
reflection.B/ c-axis pseudo cubic parameter of LVO as function of the growth temperature.}
\end{figure}

\begin{figure}[ht]
\centering \includegraphics{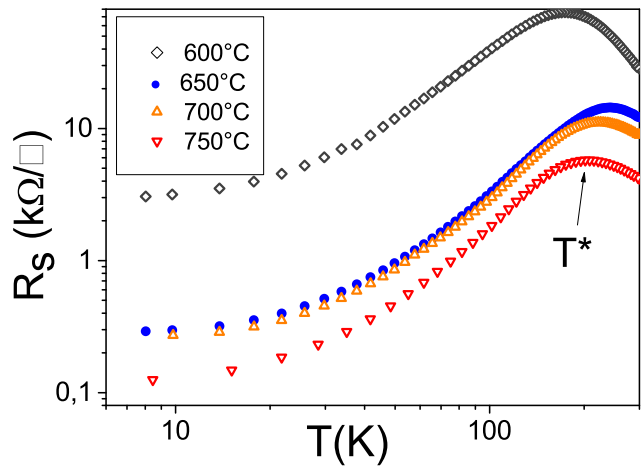}
\caption{Sheet resistance R$_{s}$ of the samples grown at different temperatures
in a log-log scale. Note the maximum of resistance at T$^{*}$.}
\end{figure}

\begin{figure}[ht]
\centering \includegraphics{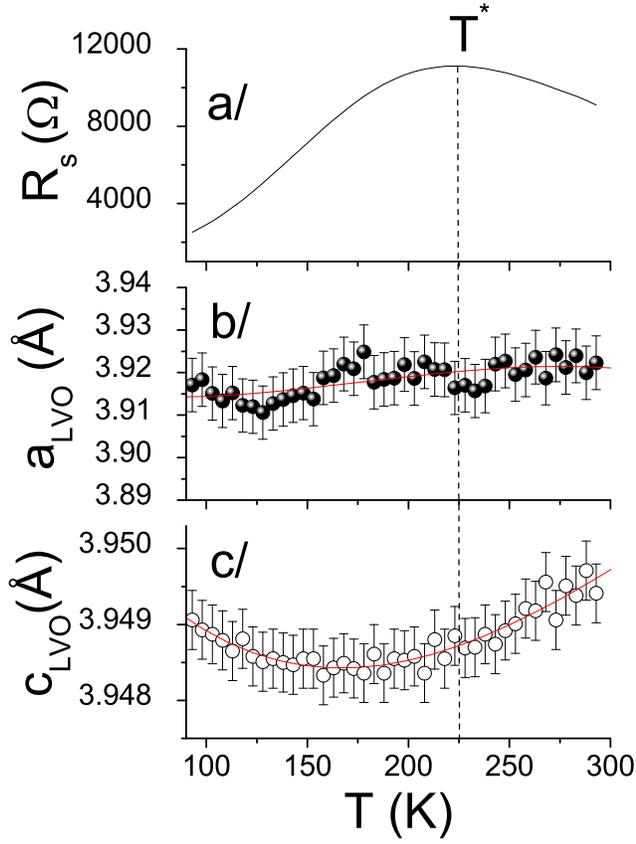}
\caption{a/ Variation of the sheet resistance of LVO/STO grown at 700$^{\circ}$ as a function of temperature.
Note the maximum at a temperature T$^{*}$signing a semiconducting-like
to metallic change of conductivity. b/ a-axis parameter of the LVO
film as a function of the temperature. c/ c-axis parameter of the
LVO film as function of temperature. Solid lines are guide for the
eyes (polynomial fits). Note the large difference between T$^{*}$
and the temperature where the c-axis parameter is minimum.}
\end{figure}

\begin{figure}[ht]
\centering \includegraphics{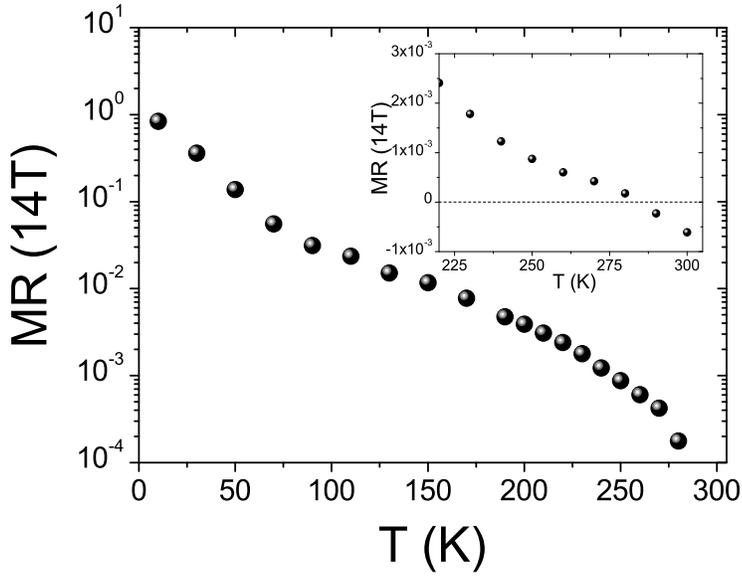} 
\caption{Magnetoresistance at 14T vs. temperature in a log-linear scale for the sample grown at 700$^{\circ}$. Note
the increase in magnitude for T $\lesssim100$K. Insert displays the cross-over between positive and negative MR at T>T$^{*}$ (linear-linear
scale).}
\end{figure}

\begin{figure}[ht]
\centering \includegraphics{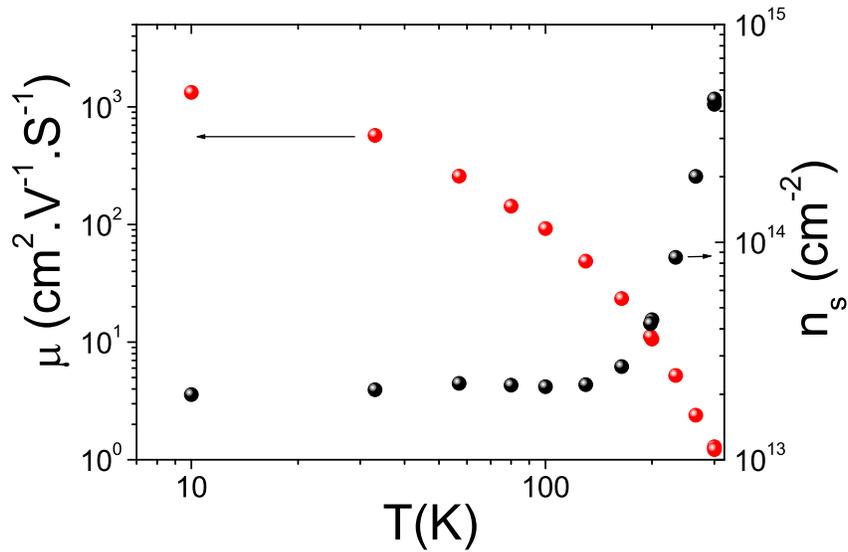}
\caption{Mobility and sheet carrier density as a function of temperature, deduced from the Hall measurements
using a single carrier model, for the sample grown at 700$^{\circ}$.}
\end{figure}

\begin{figure}[ht]
\centering \includegraphics{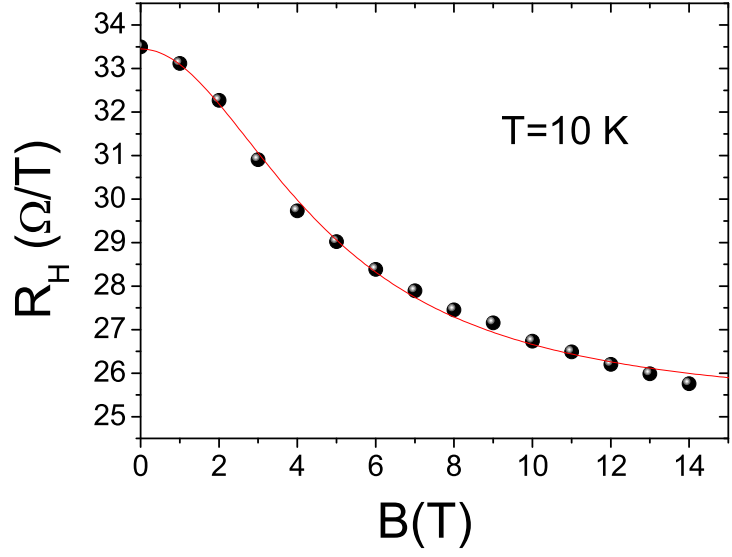}
\caption{Hall resistance R$_{H}$=R$_{xy}$/B versus the magnetic field at a temperature
T=10 K for the sample grown at 700$^{\circ}$. The solid line is a fit performed with the two-bands Hall
effect (equation (2)).}
\end{figure}

\begin{figure}[ht]
\centering \includegraphics{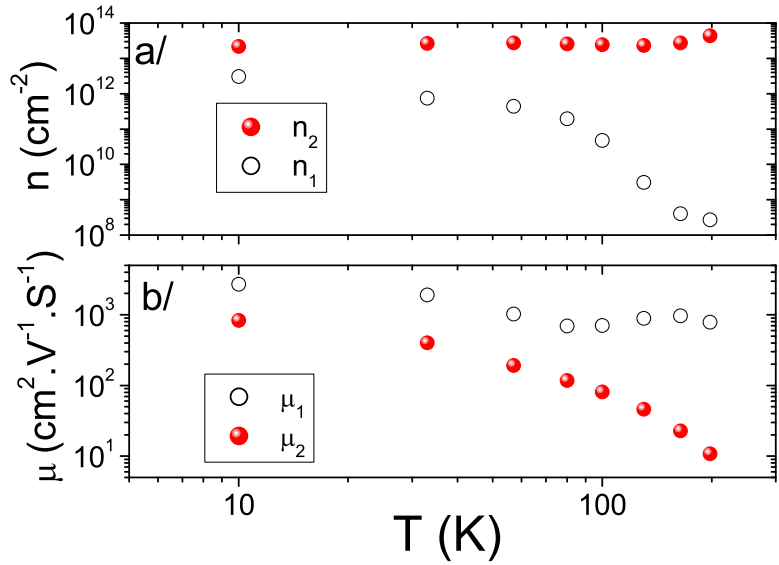} 
\caption{a/ Sheet carrier densities and b/ mobilities (bottom) of light and
heavy carriers as a function of temperature $T$, calculated using a two-bands analysis for the sample grown at 700$^{\circ}$.}
\end{figure}

\begin{figure}
\centering \includegraphics{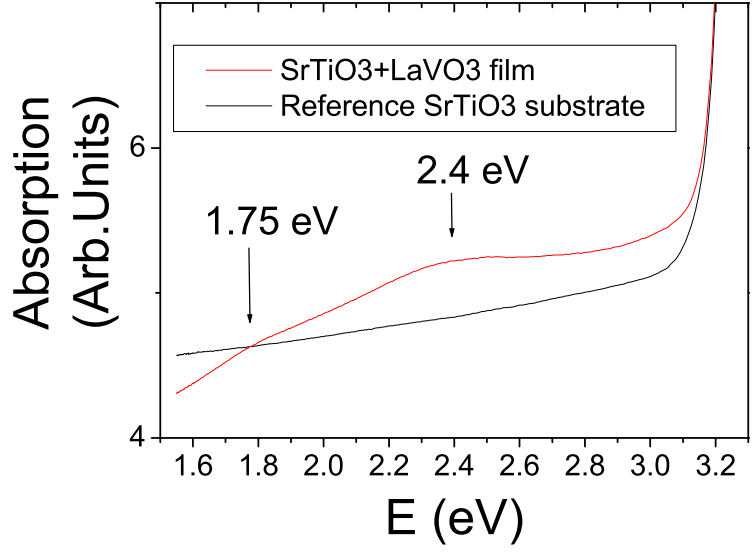}
\caption{Optical absorption for the virgin STO substrate and of (LVO/STO) grown at 700$^{\circ}$. Note the two broad peaks at 2.4 and 1.75
eV observed only for (LVO/STO), and characteristics of oxygen vacancies in STO.}
\end{figure}

\begin{figure}[ht]
\centering \includegraphics{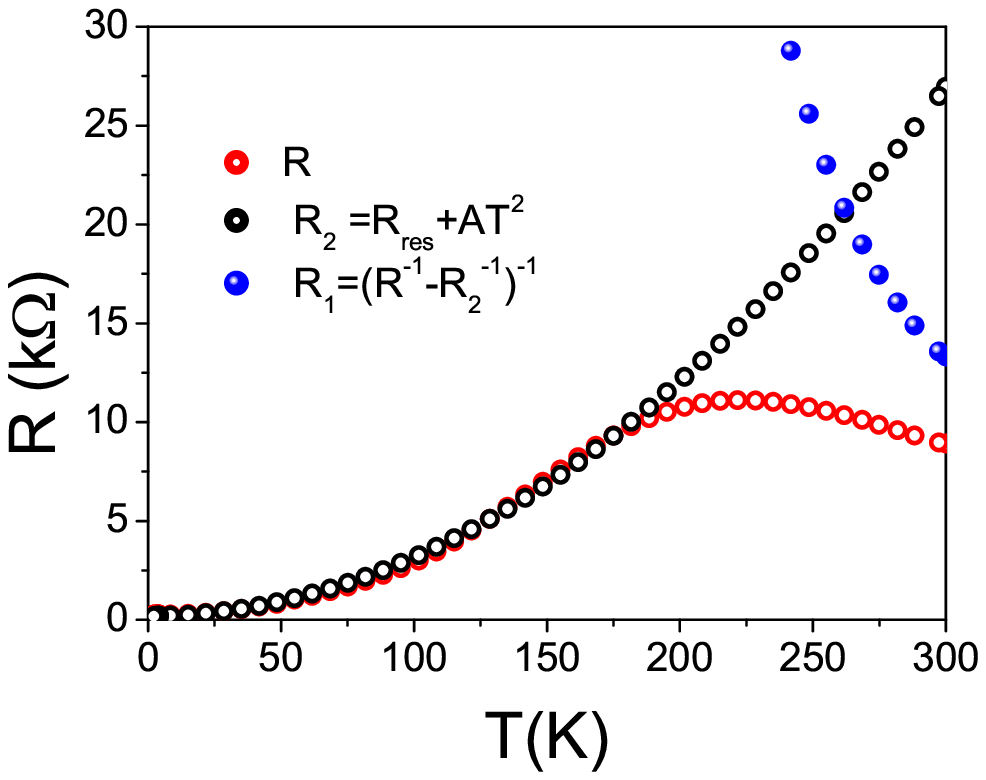}
\caption{Variation of the resistance $R$ of the sample grown at 700$^{\circ}$
as function of T temperature. Also shown R$_{2}$= R$_{res}$+AT$^{2}$
corresponding to the STO contribution and the deduced R$_{1}$ resistance
(LVO contribution) using a parallel resistor model.}
\end{figure}

\begin{figure}[ht]
\centering \includegraphics{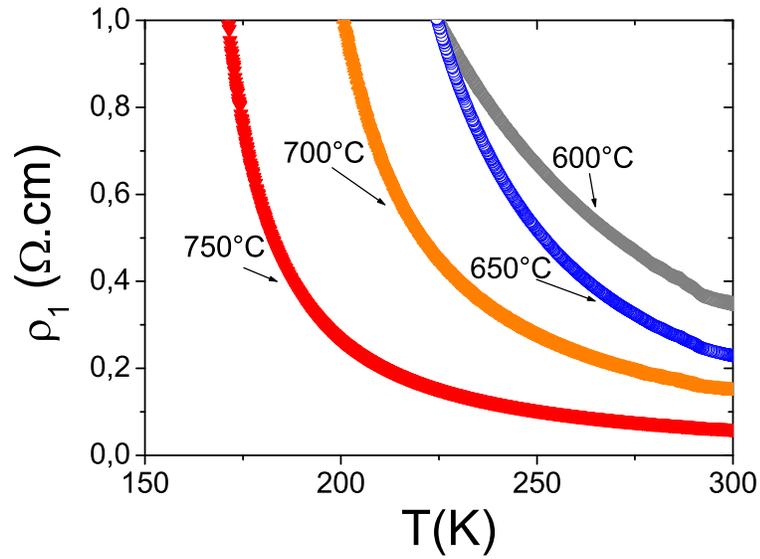}
\caption{$\rho_{1}(T)$ deduced from the parallel resistor analysis and corresponding to
the LVO component, for the different growth temperatures.}
\end{figure}

\begin{figure}
\centering \includegraphics{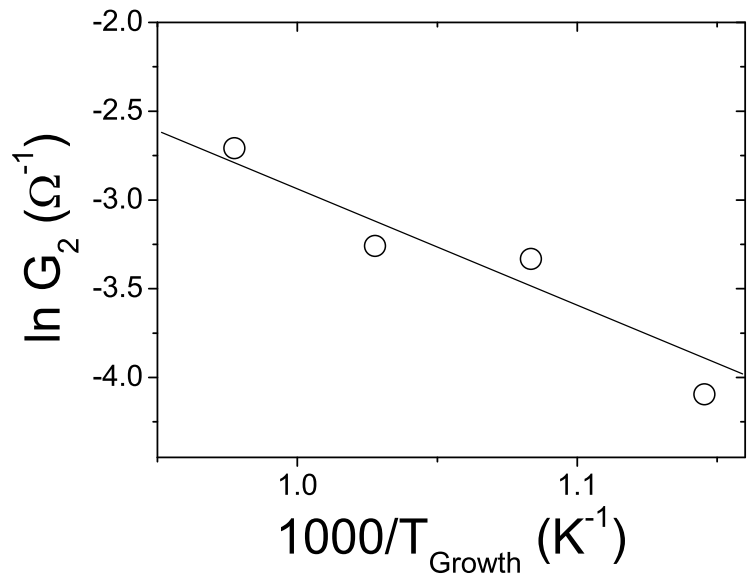}
\caption{ln G$_{2}$ as function of 1000/T$_{growth}$. The line
corresponds to an activation energy of U$\sim$0.9 eV (see text for details).}
\end{figure}

\end{document}